\documentclass[conference,letterpaper,10pt]{IEEEtran}
\usepackage{draftcopy}
\usepackage[nocompress]{cite}
\usepackage[dvips]{graphicx}\graphicspath{{./img/}}\DeclareGraphicsExtensions{.eps}
\usepackage[cmex10]{amsmath}\interdisplaylinepenalty=2500\usepackage{amssymb}
\usepackage[nearskip=0pt,farskip=0pt,captionskip=0pt,margin=0pt,font={rm},labelfont=footnotesize]{subfig}
\newcounter{tempCounter}
\newcounter{romanCounter}
\begin{document}

\title{Performance Analysis of $m$-retry BEB based DCF under Unsaturated
Traffic Condition}

\author{\IEEEauthorblockN{Md. Atiur Rahman Siddique and Joarder Kamruzzaman}\IEEEauthorblockA{Faculty of IT, Monash University, Australia\\
Email: \{Atiur.R.Siddique, Joarder.Kamruzzaman\}@infotech.monash.edu.au}}

\maketitle

\begin{abstract}
The IEEE 802.11 standard offers a cheap and promising solution for small scale wireless
networks.
Due to the self configuring nature, WLANs do not require large scale infrastructure deployment, and are scalable and easily maintainable which incited its popularity in both literature and industry.
In real environment, these networks operate mostly under unsaturated condition.
We investigate performance of such a network with $m$-retry limit BEB
based DCF.
We consider imperfect channel with provision for power capture.
Our method employs a Markov model and represents the
most common performance measures in terms of network parameters making the
model and mathematical analysis  useful in network design and planning. We also explore the effects
of packet error, network size, initial contention window, and retry limit on overall performance
of WLANs.
\end{abstract}

\section{Introduction}

Since its introduction in 1997, WLAN standard IEEE 802.11 received tremendous attention from both researchers and industry. Its widespread commercial
adoption attributes to its self configuring nature
of operation which offers easy and cheap deployment and maintenance.
In addition to providing wireless last mile coverage for the Internet, the standard has an inherent potential for integration of smart home components using cheap wireless devices.
Although wireless communication technology is also available as GSM, CDMA,
and Bluetooth, etc., designing wireless networks with these technologies imposes certain challenges for everyday use applications with high resource constraints.
For example, Bluetooth suffers from low bandwidth and short transmission range. It is attractive for personal area networks (PAN) but can not
cover large house or support community networking.
On the other hand, GSM and CDMA provide higher bandwidth and greater coverage. But the cost of installation and infrastructure maintenance makes the service cost very high rendering them inapplicable for everyday communications.
The newly introduced WiMAX, which is the industry counterpart of IEEE 802.16, overcame large scale cellular network deployment through long coverage only to offer another costly solution.
On the other hand, compared to other standards, IEEE 802.11 based WLAN offers moderate bandwidth with moderate coverage but does not require large scale infrastructure or costly devices and operates in unlicensed bands.
Due to the ad hoc nature of the standard, major
infrastructure deployment is not needed which makes it attractive for diverse
wireless applications including Ad Hoc, sensor, and mesh networks.
Only a small number of APs (Access Point) can serve Internet connectivity to a large network. Stations can join
and leave the premise of any AP without restriction.
Minimal scale infrastructure makes the operations cheap which can be availed for any kind of communication.

But the absence of infrastructure has its toll on performance.
Firstly, there is less control over the network.
Due to the inherent nature of wireless communication, new connections can
not be restricted compared to a slot based wired router.
Thus, network overloading is hard to protect against. Secondly, at each hop (closer to the AP), the number of transmissions becomes higher and a critical zone is formed around the AP limiting channel bandwidth even more.
The formation of critical zone limits the achievable throughput of
the network.
These are the reasons why a careful capacity study should be undertaken before designing and deploying such a network.

IEEE 802.11 defines two medium access methods, namely, Distributed Coordination Function (DCF) and Point Coordination Function (PCF).
DCF is more popular in literature and industry due to its ad-hoc nature of operation and inter-operability.
To arbitrate access to the channel, DCF uses a Binary Exponential Backoff (BEB) algorithm which can be either $\infty$-retry type or $m$-retry type.
If the sender keeps trying to send a packet until it succeeds, the backoff
algorithm is termed $\infty$-retry BEB where a packet can suffer repeated failures and hog the channel for long time.
$m$-retry BEB drops a packet after $m$ failures giving the next packet a
greater chance of reaching destination in time so that delay sensitive applications perform better.
In our current work, we analyze $m$-retry BEB based DCF.

A number of works studied performance of WLANs under
various conditions.
Bianchi~\cite{Bianchi2000} used a Markov model analysis
to investigate saturation throughput of a $\infty$-retry limit BEB based DCF assuming ideal channel and ignoring capture effect.
Ziouva and Antonakopoulos~\cite{Ziouva2002} presented a similar study with modified
assumptions on idle channel condition.
Bianchi's model was later extended by Wu \textit{et al.}~\cite{Wu2002} to model
$m$-retry BEB which is further extended by Chatzimisios \textit{et al.}~\cite{Chatzimisios2005,Chatzimisios2004}
to incorporate the effect of imperfect channel condition.

Although real networks operate under unsaturated load for a considerable amount of time~\cite{Daneshgaran2008}, all the above works investigated throughput under saturation condition only.
But the study of unsaturated network for both finite and infinite retry
limit BEB is of great importance in practical terms.
Barowski \textit{et al.}~\cite{Barowski2005} and Daneshgaran \textit{et al.}~\cite{Daneshgaran2008} calculated throughput under unsaturated load for an $\infty$-retry limit BEB.
Additionally, capture effect is incorporated in~\cite{Daneshgaran2008}.
Liaw \textit{et al.}~\cite{Liaw2005} presented unsaturated throughput for $m$-retry limit but the model did not consider capture effect or imperfect channel.
Most commercial 802.11 products support power capture today which brings a positive impact on overall network performance.
But to the best of the authors' knowledge, no existing work in the current literature presents an analytical model of $m$-retry BEB based DCF considering power capture.
Moreover, queuing analysis is mostly ignored which plays a crucial role in determining capacity of delay and loss sensitive applications\cite{Siddique2008,Siddique2009}.

No existing study presents a comprehensive
model and theoretical performance analyses of 802.11 WLANs considering all four factors concomitantly, namely, \setcounter{romanCounter}{1}(\roman{romanCounter})
unsaturated traffic \setcounter{romanCounter}{2}(\roman{romanCounter})~finite retry limit, \setcounter{romanCounter}{3}(\roman{romanCounter})~imperfect channel, and \setcounter{romanCounter}{4}(\roman{romanCounter})~power capture.
In this paper, we propose a simplistic Markov model considering all the
above factors and  investigate the performance of an unsaturated IEEE 802.11 compliant WLAN  in details, both theoretically and through simulation.
Consideration of imperfect channel and capture effect makes our model more suitable to reflect real world scenario closely.
Our model can accommodate both saturated and unsaturated traffic and is equally
applicable  for both basic and RTS/CTS type handshake.
Common performance measures are expressed in terms of network configuration
which makes the model useful in network design and planning.
Medium access delay and queuing loss are also estimated which will
be very useful in devising techniques to ensure QoS of delay sensitive applications
over 802.11 networks.

\section{Analytical Model}
\label{section:Mathematical Model}

We develop a Markov model for DCF mechanism employing $m$-retry BEB to analyze
performance of a homogeneous WLAN.
The model considers effects of imperfect channel and power capture and is applicable to both Basic and RTS/CTS type DCF.
It considers unsaturated traffic but can also model saturated traffic, as shown later.

\subsection{The Markov Model}
Let $n$ be the number of contending stations, $m$ the retry limit, $m'$ the number of retry stages and $p_E$ the packet error rate in an imperfect channel.
Generally, $p_E$ depends on the modulation scheme and device parameters such as transmission rate, noise, and header/data length, etc.
The model considers channel error as packet error rate and thus it can be used with any modulation scheme.
If $P\{Event\}$ is the probability of $Event$ then we define the following necessary notations and their relations for the most common probabilities
which are used in the model.
\begin{equation*}\begin{split}
\tau &{=} P\{\text{Random\ node\ transmits\ in\ random\ slot}\},\\
p_B &{=} P \{\text{Channel\ is\ busy}\}{=}1{-}(1{-}\tau)^n,\\
p_C &{=} P \{\text{Collision} | \text{Transmission}\}{=}{{(1 {-} (1 {-} \tau )^{n {-} 1} )} \mathord{\left/
 {\vphantom {{\left(1 {-} \left(1 {-} \tau \right)^{n {-} 1} \right)} {p_B }}} \right.
 \kern-\nulldelimiterspace} {p_B }}
.\\
\end{split}\end{equation*}

Capture probability $p_p$ is the probability of a collided frame being received correctly
by power capture i.e., a frame with higher power can be received even when more than one frame collided.
Assuming an 11 chip Barker sequence being used for code spreading, the spreading factor
$s$ is 11.
The receiver uses a co-relator to de-spread the original signal. The inverse of processing gain for correlation receiver, denoted by $g$, is given by
$g{=}2(3s)^{{-}1}$.
If power capture in Rayleigh fading channel is enabled, the capture probability
for a simultaneous $i$ interfering packets is given by \cite{Hadzi-velkov2002},
\begin{equation}
p_p{=}\sum_{i=1}^{n{-}1}{\binom{n}{i+1}}\tau^{i+1}(1{-}\tau)^{n-i-1}(1{+}zg)^{-i}
\end{equation}
where $z$ is the capture threshold.

$p_F$ is the probability that a transmission failed, i.e., an ACK is not received
and the packet requires a retransmission.
This can happen due to a reception with error, a collision without capture, or both.
$p_F$ can be defined as,
\begin{equation}\label{eqn:p_E}
p_F{=}(p_C{-}p_p){+}p_Ep_p{+}(p_E{-}p_Ep_C)
\end{equation}

$p_S$ is the probability that a transmitted frame is received correctly,
i.e., it is acknowledged and $p_S{=}1{-}p_F$.

We define $p_Q$ to be the probability that the queue is non-empty.
$p_Q$ is a function of the packet arrival rate $\lambda$ and channel access delay $d_C$. Assuming a $M/M/1/l_Q$ queue of length $l_Q$ with Markovian
arrival and departure, $p_Q$ can be shown to be
\begin{equation}
p_Q{=}\frac{\lambda d_C {-} (\lambda d_C)^{l_Q{+}1}}{1{-}(\lambda d_C)^{l_Q{+}1}}.
\end{equation}

Here $d_C$ is the time required to serve each packet which is given by
the interval between the time when a packet becomes the head of the queue
and the time when it is acknowledged and removed from the queue.

\begin{figure}[!t]\centering\includegraphics[width=0.33\textwidth]{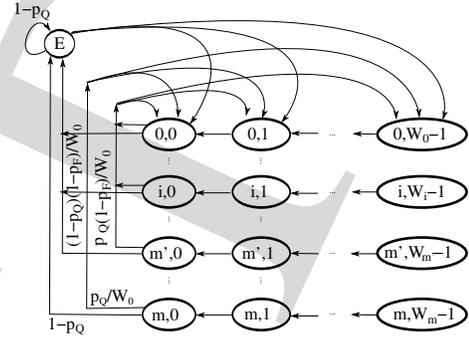}
\caption{Markov model.}\label{figure:siddi02}\end{figure}

The Markov model we used to model DCF with $m$-retry BEB is shown in Fig.~\ref{figure:siddi02}.
The states (ellipses) in a row are in the same retry stage while retry stages
increase from top to bottom.
The columns denote different counter values increasing from left to right.
The state $E$ at the top left corner of the figure represents an empty queue scenario.
This state is the key to modeling unsaturated network.
When the queue is empty, the system stays in state $E$.
Although our model resembles those presented in~\cite{Chatzimisios2004} and \cite{Daneshgaran2008}, it differs in a number of ways.
\cite{Daneshgaran2008}~modeled $\infty$-retry BEB where a packet is retransmitted
until success.
Therefore, the outgoing arcs from state ($m'$,$0$) with probability $p_F$
are recursive to $m'$-th retry stage and retry stages $m'{+}1,m'{+}2,...,m$
are not present~in\cite{Daneshgaran2008}.
Our model being $m$-retry BEB, retry stage is reset irrespective of success
and the outgoing arc from state $(m,0)$ neither is recursive,
nor depends on success or failure of the last transmission.
On the other hand, \cite{Chatzimisios2004} modeled saturation traffic.
Therefore, when the retry stage is reset, the Markov model always goes to the $0$-th retry stage and empty queue state $E$ is not present.
In our model, if a transmission is successful or the retry limit is exceeded, $E$ becomes the new state if the queue is empty.
Although \cite{Liaw2005} investigated unsaturated load for $m$-retry BEB,
the effects of imperfect channel and power capture are ignored.

We define the Markov model states as a bi-dimensional process $\left\{s(t),b(t)\right\}$  where $s(t)$ is the retry stage and
$b(t)$ is the counter value at time $t$.
The current backoff window $W_i$ at retry stage $i$ is defined as,
\begin{equation*}
W_i{=}\begin{cases}
2^iW
&\text{for } 0\leqslant i\leqslant m',\\
2^{m'}W
&\text{for } i>m'.\\
\end{cases}\end{equation*}

\subsection{One Step Transition Probabilities}
We denote the one step transition probability  from state $(i,k)$ to $(i',k')$ with $P\{i',k'|i,k\}$ are defined as,

\begin{equation*}\setlength{\arraycolsep}{0pt}
\begin{array}{l}
P\{i,k|i,k{+}1\}{=}1  \text{\ for } \ 0{\leqslant} i{\leqslant} m\text{ and }1{\leqslant} k{\leqslant} W_{i}{-}2,\\
P\{i,k|i{-}1,0\}{=}
\frac{p_F}{W_i}  \text{ for } 1{\leqslant} i{\leqslant} m\text{ and }\ 0{\leqslant} k{\leqslant} W_i{-}1,\\
P\{0,k|i,0\}{=}\frac{p_Q(1{-}p_F)}{W}\text{ for } 0{\leqslant} i{\leqslant}  m{-}1,0{\leqslant} k{\leqslant} W{-}1,\\
P\{0,k|m,0\}{=}\frac{p_Q}{W} \text{\ for\ } 0{\leqslant} k{\leqslant} W{-}1,\\
P\{E|i,0\}=(1-p_Q)(1-p_F) \text{\ for\ } \ 0{\leqslant} i{\leqslant} m{-}1,\\
P\{E|m,0\}{=}(1-p_Q) \text{ for } 0{\leqslant} i{\leqslant} m{-}1,\\
P\{0,k|E\}{=}\frac{1}{W} \text{\ for\ } 0{\leqslant} k{\leqslant} W{-}1.
\end{array}
\end{equation*}

\subsection{Stationary Probabilities of the Markov States}

We define $b_{i,k}$ as the stationary probability distribution of any state $(i,k)$
which is given by,
\begin{equation*}
b_{i,k}{=}\lim\limits_{t\to \infty}P\{s(t){=}i,c(t){=}k\},i\in[0,m],k\in[0,2^iW].
\end{equation*}

Let the steady state probability of the system being in state $E$ be denoted by $b_E$.
As long as
the retry limit is not exceeded, the retry stage increments after each failed
transmission and
\begin{equation*}
b_{i,0}{=}b_{i{-}1,0}p_{F}{=}p_{F}^ib_{0,0} \text{ for } 1{\leqslant} i{\leqslant} m .
\end{equation*}

Due to chain regularity,
\begin{equation*}
b_{i,k}{=}\frac{W_i{-}k}{W_i}
\begin{cases}
p_Q(1{-}p_{F})\sum\limits^{m{-}1}_{j{=}0}b_{j,0}{+}p_{Q}b_{m,0}
& \text{for } i{=}0,\\
p_{F}b_{i{-}1,0}
&\text{for } 1{\leqslant} i{\leqslant} m.\\
\end{cases}\end{equation*}

The steady state probability of entering and leaving any state is equal.
Imposing this condition on state $E$ we obtain,
\begin{equation}
\label{equation:b_E}
\setlength{\arraycolsep}{0pt}
b_E{=}\frac{1}{p_Q}\left\{(1{-}p_Q)(1{-}p_{F})\sum\limits_{i{=}0}^{m{-}1}b_{i,0}{+}(1{-}p_Q)b_{m,0}\right\}.
\end{equation}
Imposing the same condition on state $\{0,0\}$ gives,
\begin{equation}\label{equation:b_E,simple}
p_{F}b_{0,0}
{=}(1{-}p_{F})\sum\limits_{i{=}0}^{m}b_{i,0}{+}p_{F}b_{m,0}.
\end{equation}
Sum of all states, where a packet is transmitted, gives,
\begin{equation}
\label{equation:sum of b_i,0}
\sum\limits_{i{=}0}^{m{-}1}b_{i,0}{=}b_{0,0}\sum\limits_{i{=}0}^{m{-}1}p^{i}{=}\frac{1{-}p^{m}}{1{-}p_F}b_{0,0}.
\end{equation}
From (\ref{equation:b_E}) \& (\ref{equation:sum of b_i,0}), we derive
\begin{equation*}
b_E
 {=}\frac{1{-}p_Q}{p_Q}b_{0,0}.\\
\end{equation*}
\begin{figure*}[!t]
\normalsize
\setcounter{tempCounter}{\value{equation}}
\setcounter{equation}{7}
\begin{equation}
\label{equation:b_{0,0}}
b_{0,0}{=}\frac{2p_Q(1{-}p_F)(1{-}2p_F)}{
\begin{split}
p_QW(1{-}p_F)\left(1{-}(2p_F)^{m'{+}1}\right){+}p_Q(1{-}2p_F)(1{{-}}p_F^{m'{+}1}){+}p_Q
p_F^{m'{+}1}(1{-}2p_F)(2^{m'}W{+}1)(1{-}p_F^{m{-}m'})\\
\:{+}2(1{-}p_F)(1{-}2p_F)(1{-}p_Q)(1{-}p_F^m){+}2p_F^m(1{-}p_F)(1{-}2p_F)(1{-}p_Q)
\end{split}}.
\end{equation}
\setcounter{equation}{\value{tempCounter}}
\hrulefill
\end{figure*}
The  sum of probabilities of being in every state should be equal to $1$ i.e.,$\sum\limits_{i{=}0}^{m}\sum\limits_{k{=}0}^{W_i{-}1}{b_{i,k}}{+}b_E{=}1
$ which gives (\ref{equation:b_{0,0}}).
Under saturation condition, a node will always have some packet to send.
Therefore, $p_Q{\to}1$
which gives $b_{0,0}$ for saturation model as shown in~\cite{Chatzimisios2004}.
A packet is transmitted only from the states $b_{i,0}$ where $i{\in}[0,m]$. Since $\tau$ denotes
the probability of transmission by a random node at a random slot,
\setcounter{equation}{8}
\begin{equation}
\label{eqn:tau}
\tau{=}\sum\limits_{i{=}0}^{m}b_{i,0}{=}b_{0,0}\sum\limits_{i{=}0}^{m}p_F^{i}{=}\frac{1{-}p_F^{m{+}1}}{1{-}p_F}b_{0,0}.
\end{equation}

Let $t_{\sigma}$, $t_{sifs}$, and $t_{difs}$ be the length of an idle slot, SIFS, and DIFS period.
$t_{\delta}$ is the propagation delay.
Similarly, $t_D$, $t_A$, $t_R$, and $t_C$ are the transmission time of data frame, ACK frame, RTS frame, and CTS frame, respectively.

A successful transmission is detected by the reception of an ACK packet.
We define $t_S$ as the time a sender has to wait before receiving an ACK
for the last packet and remove it from the queue. Taking into account of all the delay elements, $t_S$ can be defined as
\begin{equation*}\setlength{\arraycolsep}{0pt}
t_S {=}
\begin{cases}
t_{difs} {+} t_D {+} t_{\delta} {+} t_{sifs} {+} t_A {+} t_{\delta} 
&\text{for Basic,}\\
\begin{split}
t_{difs} {+} t_R {+} t_{\delta} {+} t_{sifs} {+} t_{C} {+} t_{\delta} {+}
t_{sifs}\\
{+} t_D {+} t_{\delta} {+} t_{sifs} {+} t_A {+} t_{\delta}\\
\end{split} &
\text{for RTS/CTS}.\\
\end{cases}
\end{equation*}
When a transmission fails, the sender can detect neither collision nor erroneous
reception.
Therefore, it has to wait for a time duration $t_F$  before taking the transmission to be failed, which can be defined as
\begin{equation*}\setlength{\arraycolsep}{0pt}
t_F {=}\begin{cases}
t_{difs} {+} t_D {+} t_{\delta} {+} t_{sifs} {+} t_A {+} t_{\delta}
&\text{for Basic,}\\
t_{difs} {+} t_R {+} t_{\delta} {+} t_{sifs} {+} t_{C} {+} t_{\delta}
&\text{for RTS/CTS.}\\
\end{cases}
\end{equation*}
Using $p_F$ and $\tau$ defined as in (\ref{eqn:p_E}) and (\ref{eqn:tau}),
respectively, we calculate the expected slot length $t_{slot}$ as,
\begin{equation*}\setlength{\arraycolsep}{0pt}
t_{slot} {=}(1 {-} p_B) t_{\sigma} {+} p_B (1 {-} p_S) t_F {+} p_B p_S p_E t_F {+} p_B  p_S  (1 {-} p_E)  t_S.
\end{equation*}

\begin{figure*}\centering
\subfloat[Effect of payload, $l_D$]{\includegraphics[width=0.33\textwidth]{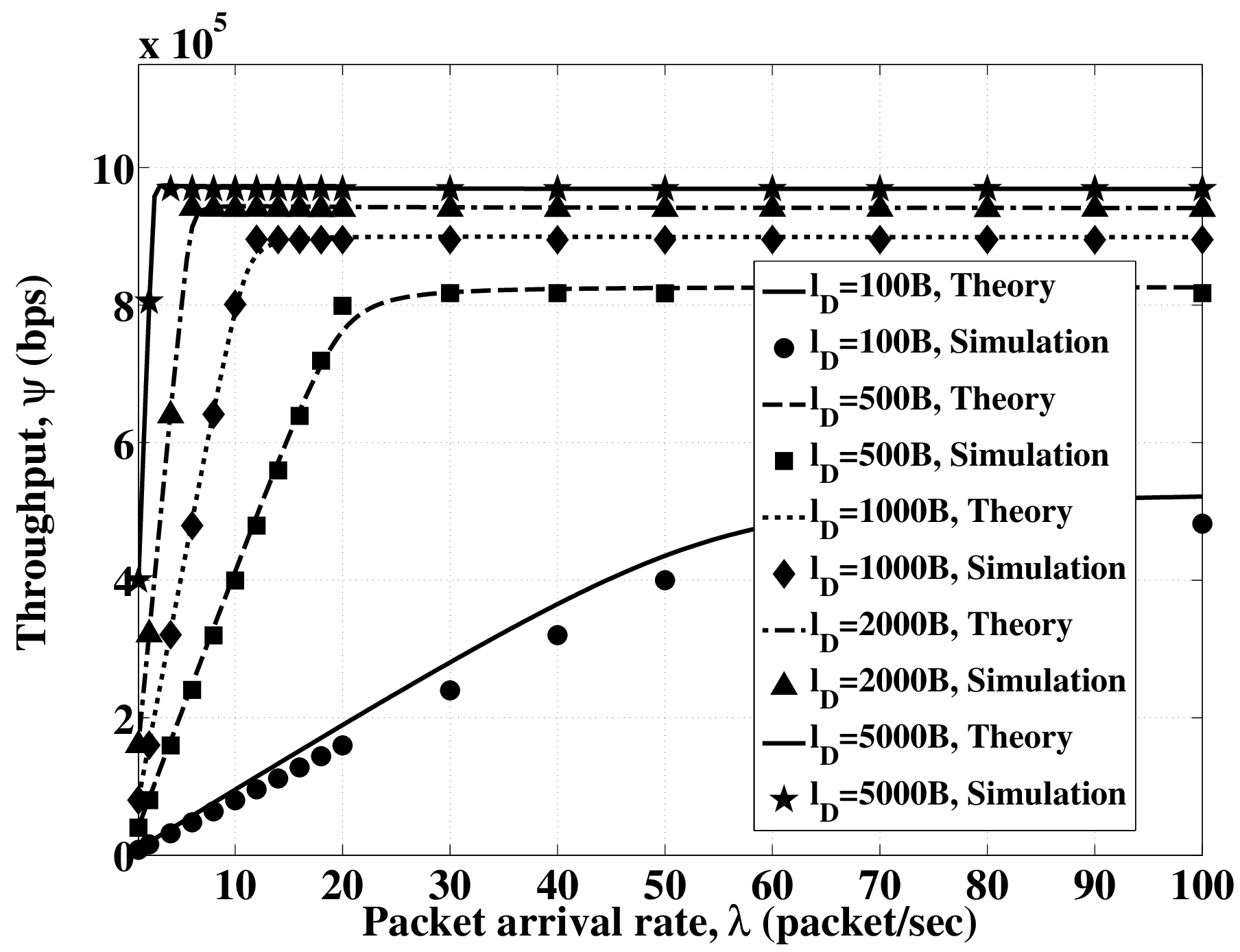}\label{fig:packetSize}}
\hfill
\subfloat[Effect of packet error rate, $p_E$]{\includegraphics[width=0.33\textwidth]{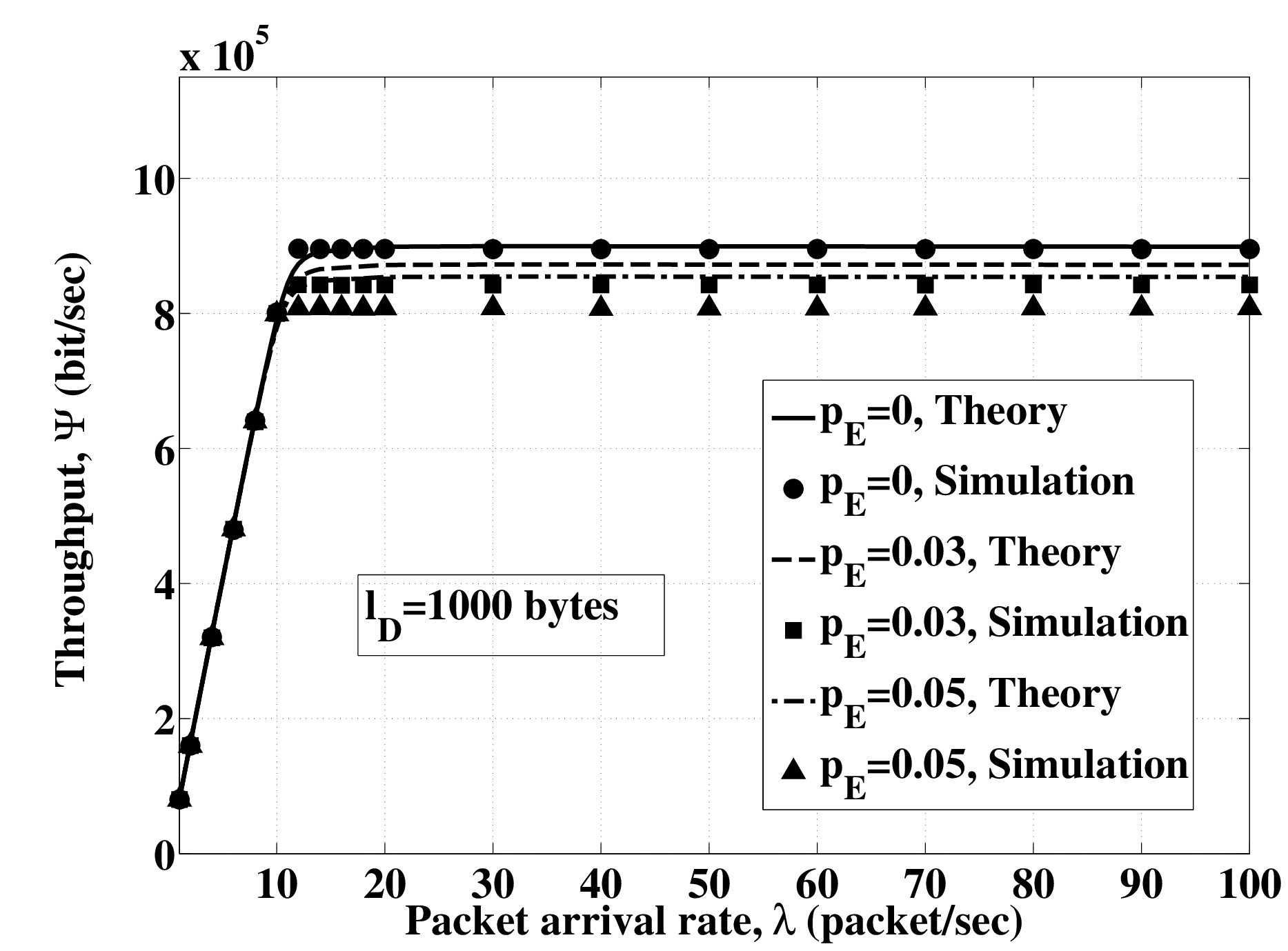}\label{fig:P_e-2}}
\hfill
\subfloat[Effect of number of stations, $n$]{\includegraphics[width=0.33\textwidth]{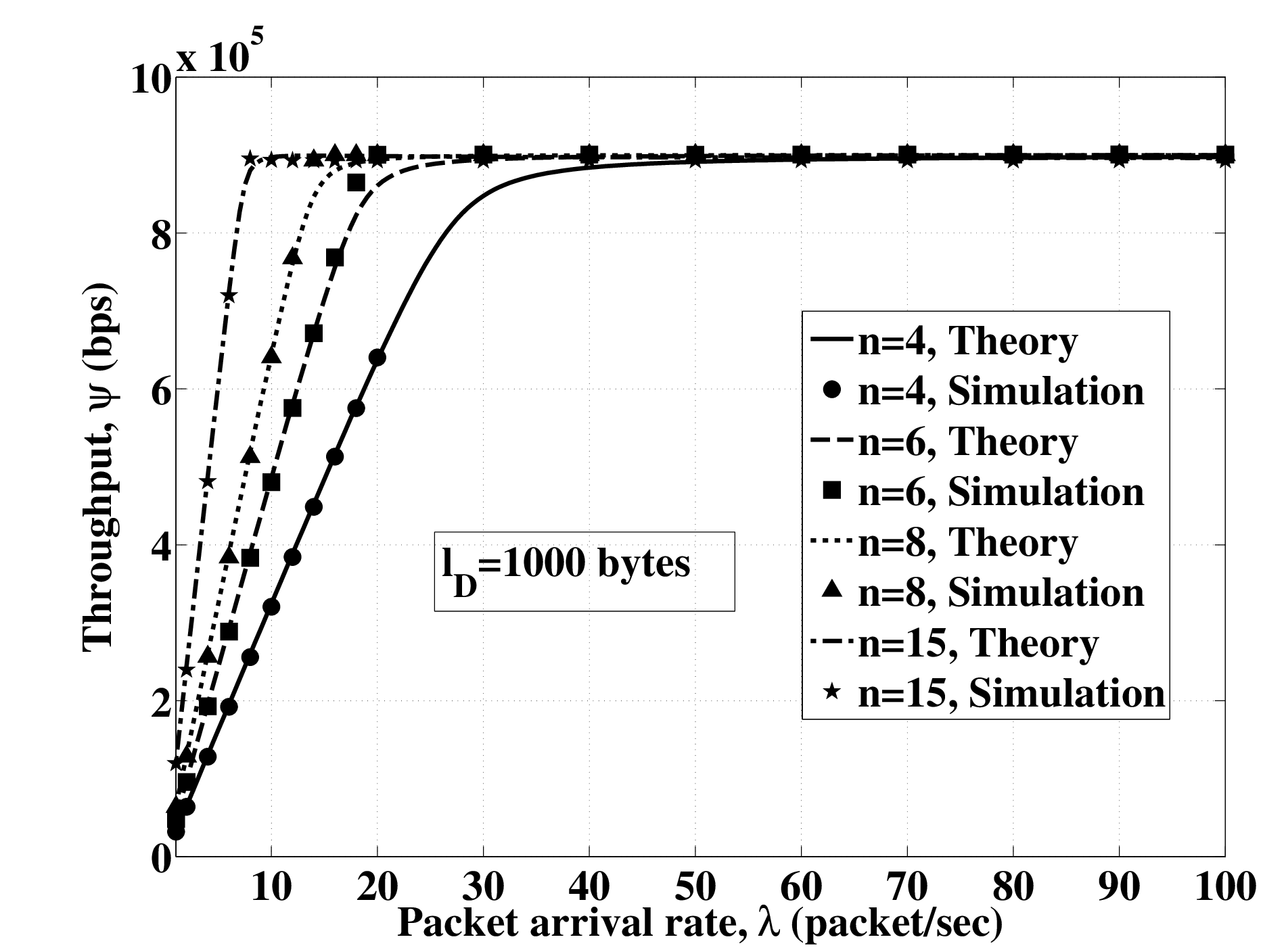}\label{fig:nn}}
\caption{Simulation and analytical throughput for different (a) payload, (b) packet error rate,
and (c) number of stations.}
\label{fig:sim}
\end{figure*}

\subsection{Performance Measures}
Throughput $\psi$ is the number of payload bits that the MAC layer can transmit per second which is given by,
\begin{equation*}
\psi{=}\frac{p_Bp_S(l_D+l_I+l_U)}{t_{Slot}}.\\
\end{equation*}
where $l_D$, $l_I$, and $l_U$ are length of application data, IP header, and UDP/TCP header.

A packet is dropped when it suffers from more than $m$ number of failed
transmissions. Therefore, network packet loss $e_N$ is the probability that
a packet suffers from at least $m{+}1$ failed transmissions and is given by~\cite{Wu2002},
\begin{equation*}
e_N {=} p_F^{m{+}1}.
\end{equation*}

If queue length is not very large, queuing loss $e_Q$ can play an important
role for loss sensitive applications. For the $M/M/1/l_Q$ queue described earlier, $e_Q$ is given by,
\begin{equation*}
e_Q{=}\begin{cases}
\frac{(\lambda d_C)^{l_Q}{-}(\lambda d_C)^{l_Q{+}1}}{1-(\lambda d_C)^{l_Q{+}1}}&\text{when } \lambda d_C {\neq} 1,\\ 
\frac{1}{l_Q{+}1} 
& \text{otherwise.}
\end{cases}
\end{equation*}

Channel access delay $d_C$ is defined to be the length of the period starting when a packet becomes the head of the queue (HoQ) and ending when an ACK frame
confirms its successful reception.
Delay faced by the dropped packets (because of exceeding retry limit) does not
contribute to it.
The probability that a packet is successfully transmitted is $1{-}e_N{=}1{-}p_F^{m+1}$.
A packet may face delay at each retry stage and the probability that it reaches
stage $i$ and is not dropped (i.e., it is successful) is given by $\frac{p_F^i{-}p_F^{m{+}1}}{1{-}p_F^{m{+}1}}$.
At retry stage $i$, a packet can face $\frac{W_i}{2}$ number of slots on average.
In total, the packet has to wait for $n_{slot}$ number of slots as HoQ which is given by,
\begin{equation*}
n_{slot} {=} \sum_{i{=}0}^{m}\frac{W_i {+} 1}{2} \frac{p_F^i{-}p_F^{m{+}1}}{1{-}p_F^{m+1}}.
\end{equation*}

Using the above expression for number of slots, the channel access delay can
be estimated as~\cite{Wu2002},
\begin{equation*}
d_C=n_{slot}t_{slot}.
\end{equation*}

\section{Simulation and Results}
\label{section:Simulation and Results}
We performed all mathematical analyses in Matlab 2007a and
modeled the simulations in Network Simulator 2 (NS-2.28) which is widely
 used by network researchers.
We found a very agreeing match, as shown later, between the simulations and our theoretical datasets which validates our model.
NS-2's original tracing mechanism being slow and inadequate to trace a packet
at each network layer, a separate tracing mechanism is developed and used in this work.
To carefully study the effect of network load and packet arrival rate, a
separate Poisson agent is also developed along with a corresponding AP (Null) agent.
These classes were made open source and posted in NS-2 official mail group for comments and public use.

We tested the simulations for IEEE 802.11B with DSSS based physical layer but the method is applicable
to any 802.11 variant.
The parameters used in the simulation are shown in Table~\ref{table:simulation parameters}.
\begin{table}[!t]\renewcommand{\arraystretch}{1.3}\caption{Simulation parameters}\label{table:simulation parameters}\centering
\begin{tabular}{|c|c|c|c|}
\hline
Parameter & Value &\ Parameter & Value\\
\hline
SIFS & $10\mu s$ &
Idle slot & $20\mu s$\\
DIFS & $50\mu s$ &
Propagation delay & $1\mu s$\\
PLCP header & $144b$ &
Preamble & $48b$\\
Data rate & $1$ Mbps&
Basic rate & $1$ Mbps\\
PLCP rate & $1$ Mbps&
Run length& 2000s \\
\hline
\end{tabular}
\end{table}
Fig.~\ref{fig:sim}\subref{fig:packetSize} shows throughput $\psi$ as a function of packet arrival rate $\lambda$ for different payload sizes.
We used a wide range of payload sizes ($100{\sim}5000$ bytes).
The simulation results match mathematical data closely.
The throughput increases almost linearly with increasing arrival
rate up to a certain point, beyond which it does not increase any further.
The first region indicates unsaturated condition of the network where $\lambda {\ll} \frac{1}{d_C}$ while the second region indicates near saturation or saturation region where $\lambda {\geq} \frac{1}{d_C}$.
The smooth transition between the two regions is where network switches into saturation and $\lambda{\approx}\frac{1}{d_C}$ (e.g., $\lambda{=}20{\sim}30$ packets/sec for $l_D$=500B).
This transition is clearly a function of the payload and it is achieved at a lower $\lambda$ when the payload is larger.

\begin{figure*}\centering
\subfloat[Effect of initial collision window, $W$]{\includegraphics[width=0.33\textwidth]{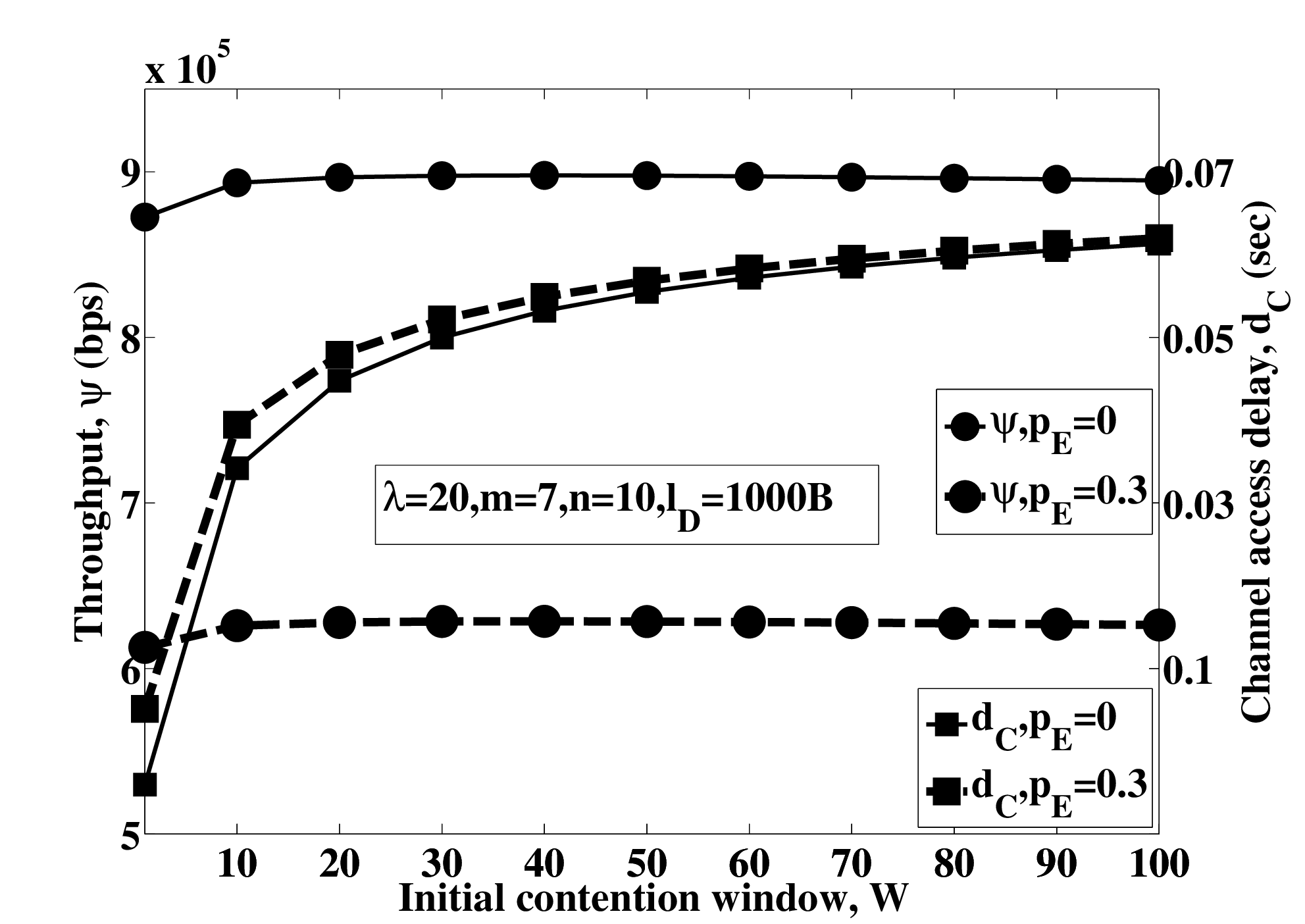}\label{fig:A1}}
\hfill
\subfloat[Effect of retry limit, $m$] {\includegraphics[width=0.33\textwidth]{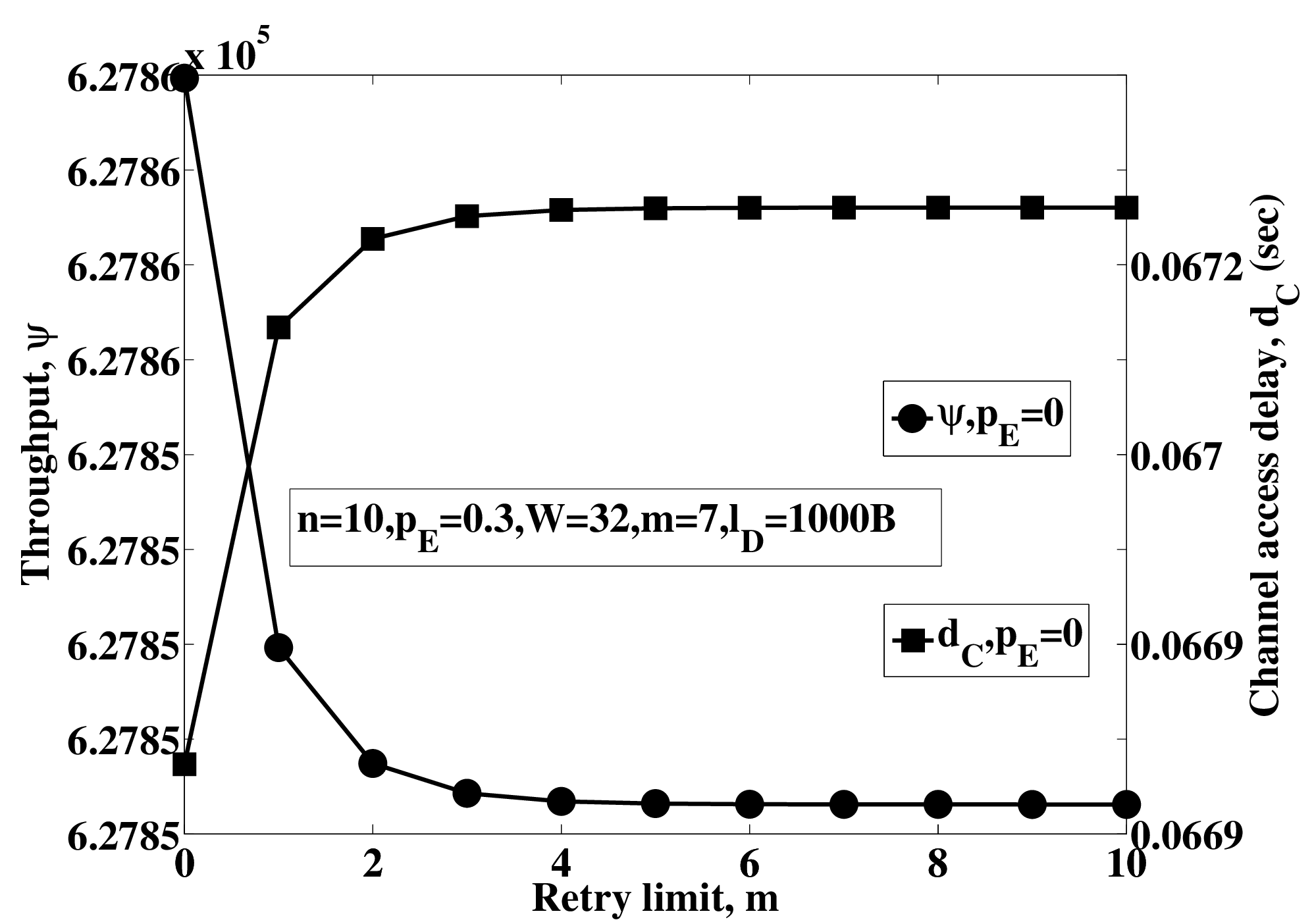}\label{fig:A3A}}
\hfill
\subfloat[Effect of number of stations, $n$]{\includegraphics[width=0.33\textwidth]{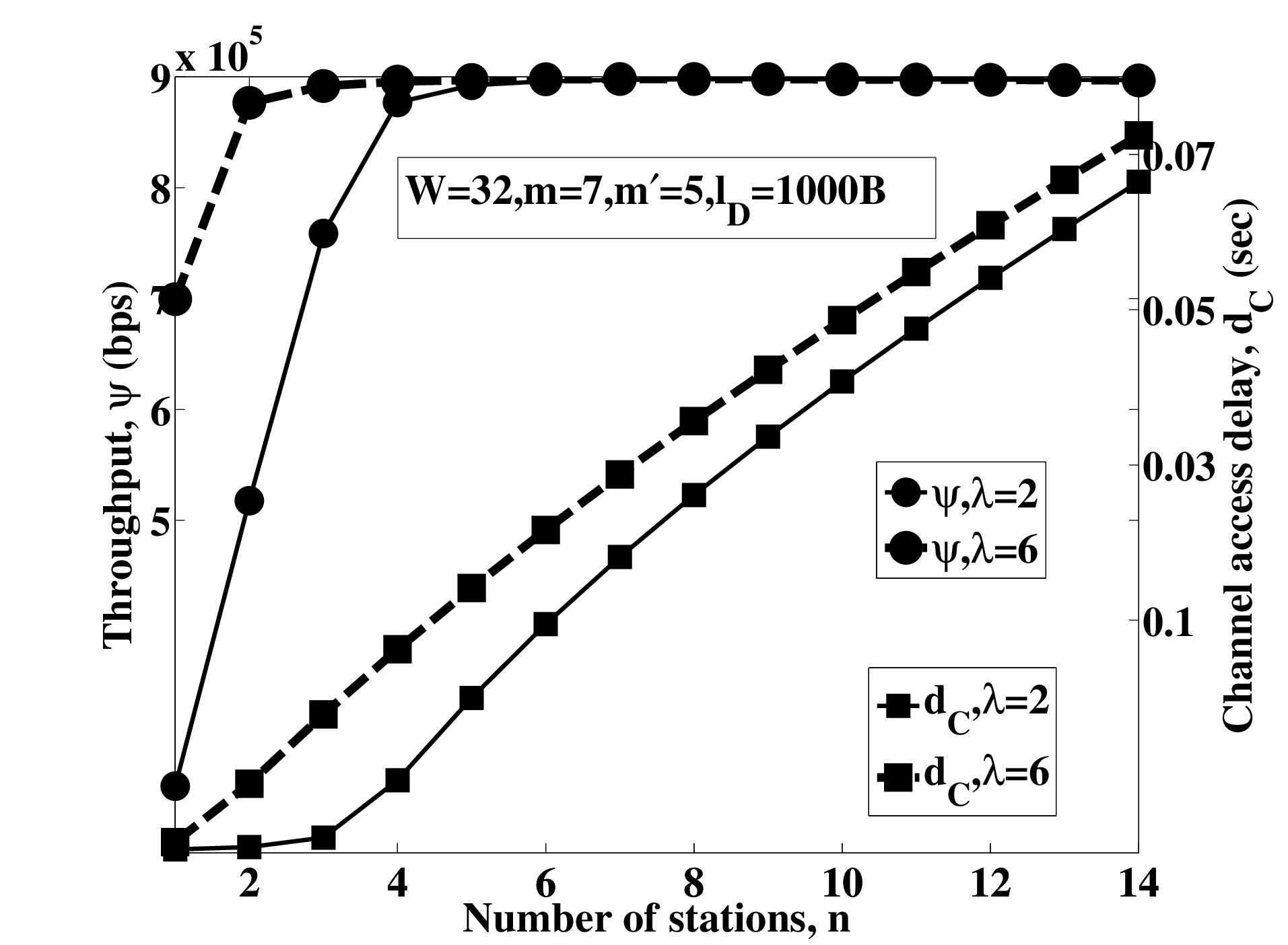}\label{fig:A4A}}
\caption{Effect of (a) initial collision window, (b) retry limit, and (c) number
of stations on throughput and delay.}
\label{fig:math}
\end{figure*}

\begin{figure*}\centering
\subfloat[Effect of packet arrival rate, $\lambda$]{\includegraphics[width=0.33\textwidth]{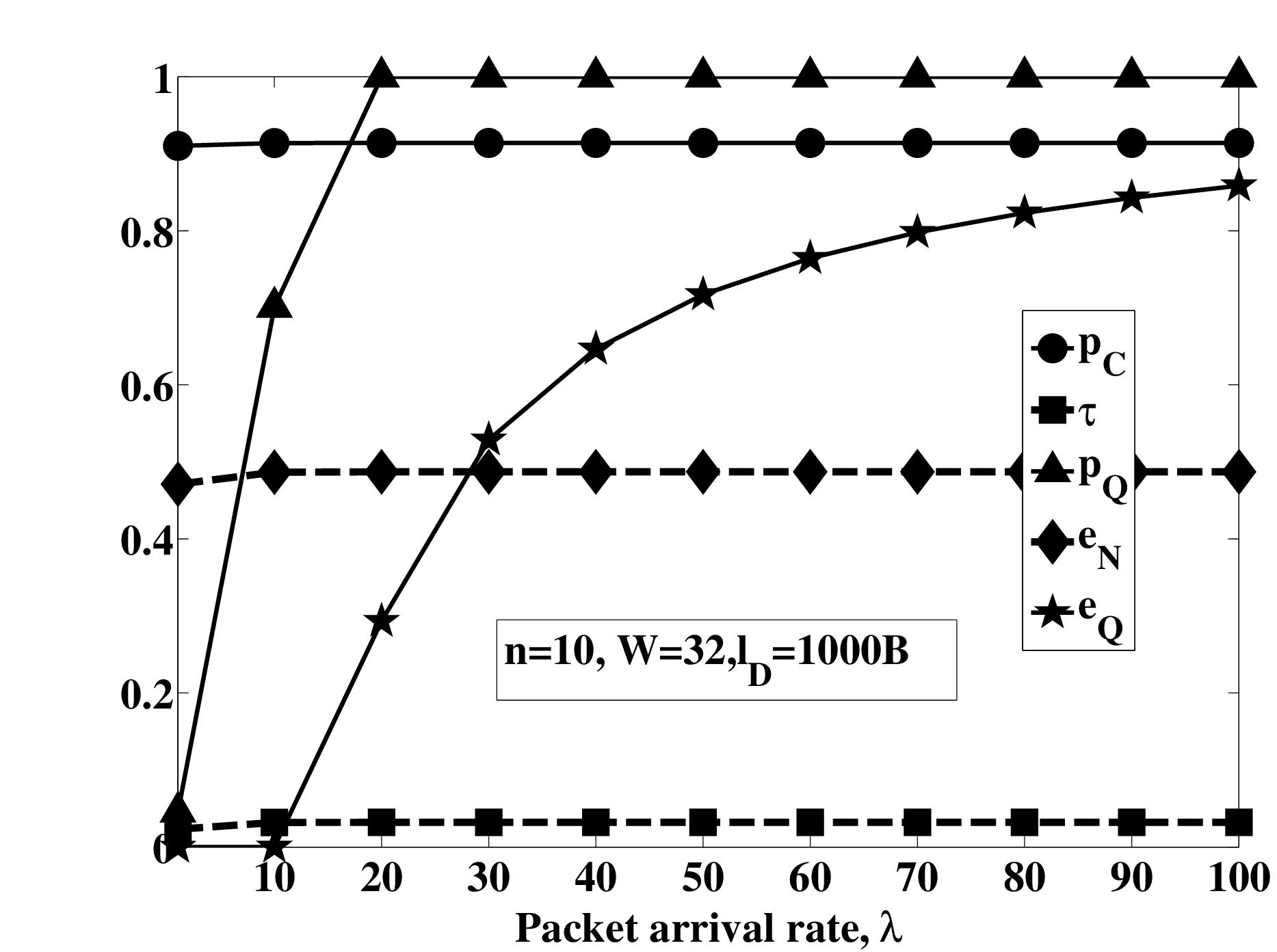}\label{fig:B1}}
\hfill
\subfloat[Effect of initial collision window, $W$]{\includegraphics[width=0.33\textwidth]{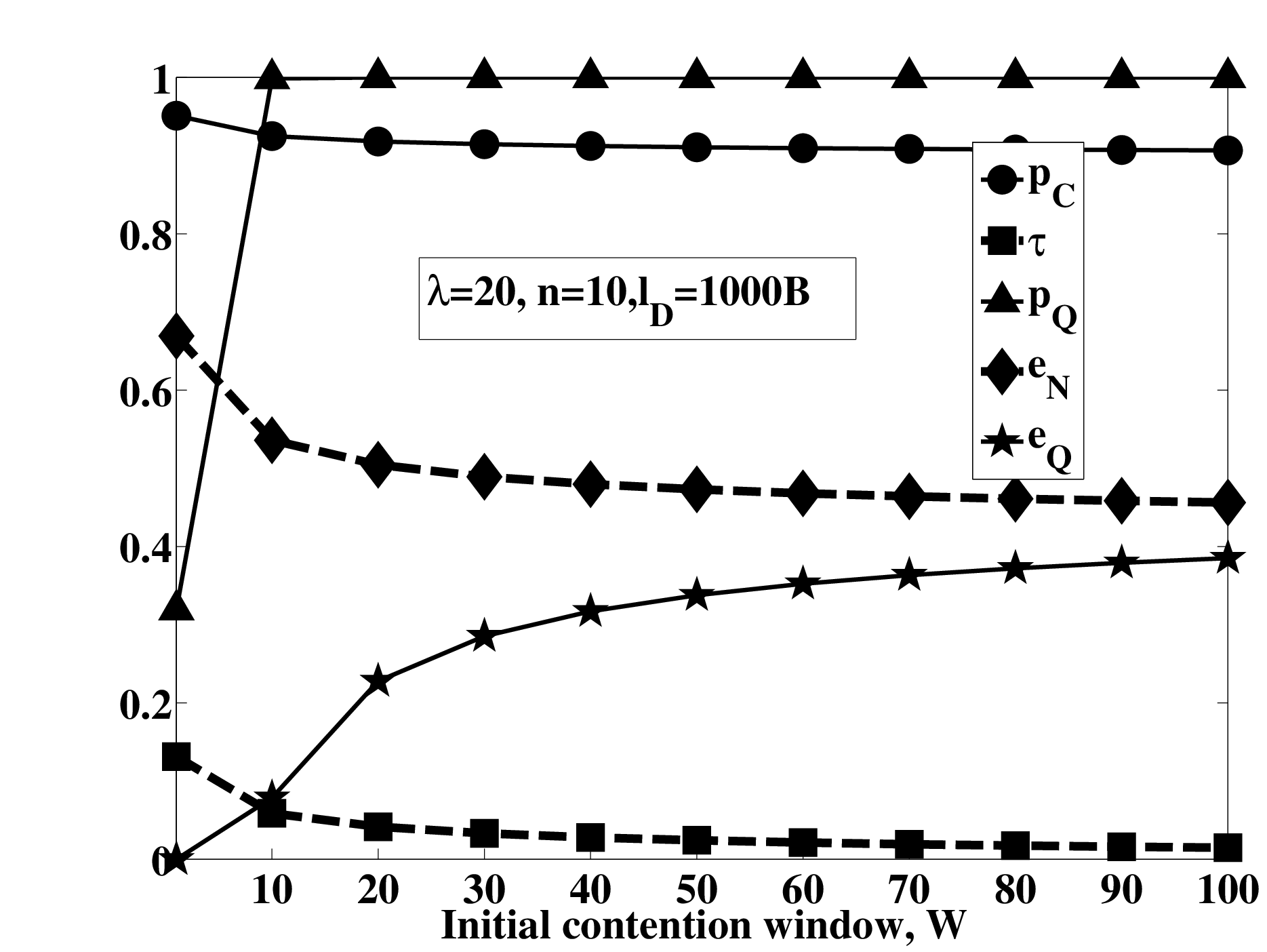}\label{fig:B4}}
\hfill
\subfloat[Effect of retry limit, $m$]{\includegraphics[width=0.33\textwidth]{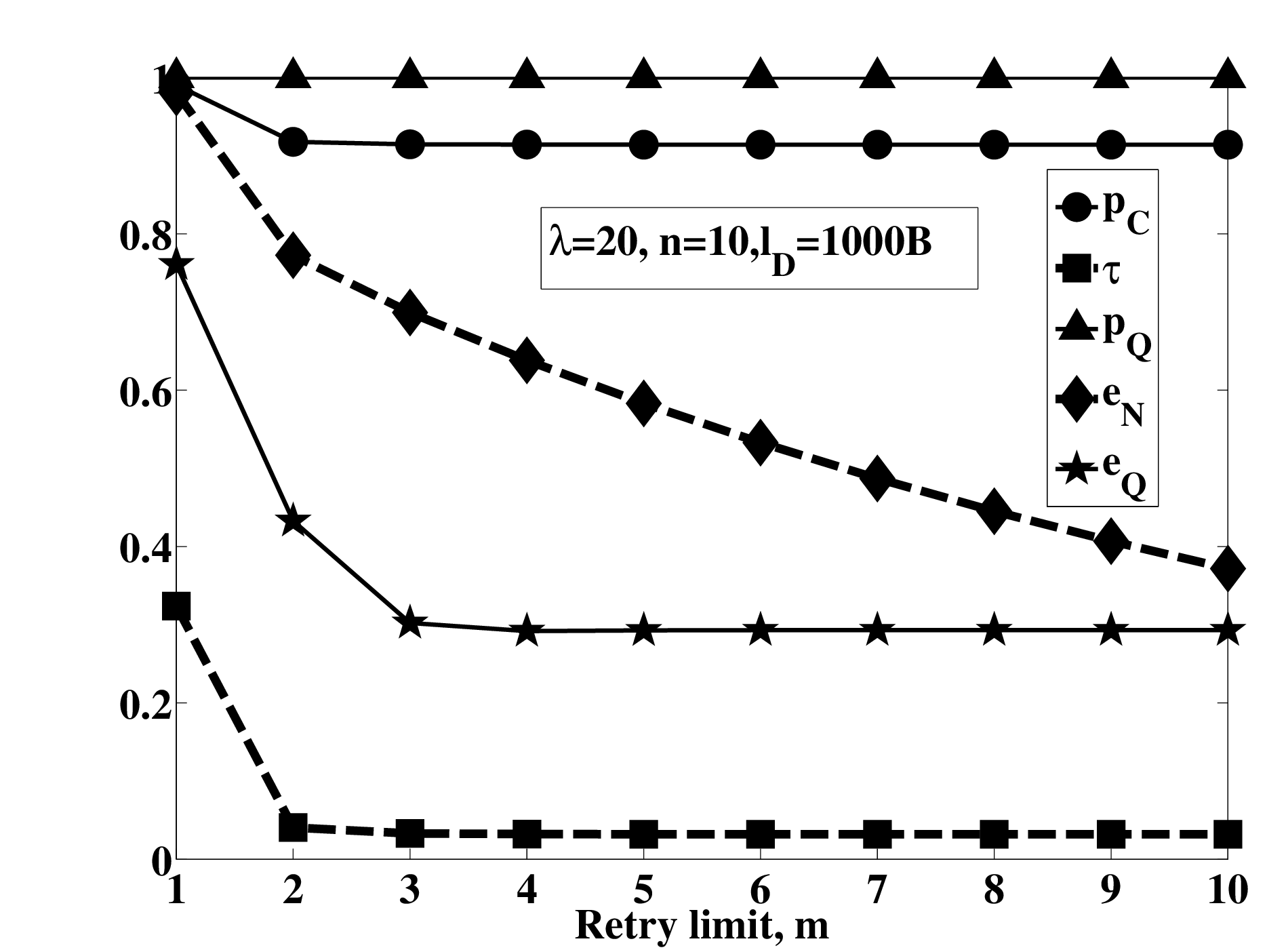}\label{fig:B7}}
\caption{Effect of a) arrival rate, b) initial contention window, and c) retry limit on probability of collision $p_C$, probability of transmission $\tau$, probability of non-empty queue
$p_Q$, network loss $e_N$, and queuing loss $e_Q$. }
\label{fig:queue}
\end{figure*}

For a larger payload, the transmission time becomes longer, making the transmissions more susceptible to collisions and errors.
Thus the channel access time becomes longer saturating the network even with a smaller packet arrival rate.
A longer channel access time incurs relatively high degradation in throughput since the channel is kept busy for each packet in the queue.
However, the payload contributes even more to the total throughput and a greater payload still gives a greater throughput although channel access time is longer in this case.
This is the reason the rate of increase of throughput in the unsaturated region is higher for longer packets and the same level of throughput can be achieved by longer packets at a lower packet arrival rate.
For the rest of this study, we use $l_D=1000$ bytes as a representative data
payload size.

Fig.~\ref{fig:sim}\subref{fig:P_e-2} shows change in throughput for different packet error rates.
These results show that although the impact of packet error rate is considerably high in saturated region, its impact is negligible in unsaturated region.
In simulation, errors are introduced uniformly in the received packets.
Since the number of transmissions in the unsaturated region is quite low, less number of retries is necessary and hence the  impact is found insignificant.
On the other hand, in the saturated region, a large number of transmissions and retransmissions take place; as a result the impact of erroneous packets on overall throughput becomes greater.

Fig.~\ref{fig:sim}\subref{fig:nn} presents results for different numbers of stations in the network.
Since the payload is kept constant to 1000 bytes, the saturation throughput is same in all cases which is 0.9 Mbps.
Below the saturation point, i.e., in the unsaturated region, the rate of increase in throughput is different for different network sizes.
Both the packet arrival rate and the number of nodes define the load on the network since the number of generated packets depends on them.
The number of nodes also means the number of competitors for the channel.
Therefore, even if the number of generated packets in unit time (and payload) is same, a greater number of nodes may lead to a greater number of collisions, which in turn will result in a higher channel access time and a lower throughput.
The results demonstrated the same phenomenon where, every thing else being unchanged, a higher number of nodes achieved saturation even with a much lower packet arrival rate.
While a $4$-node network did not achieve saturation until $\lambda{=}60$, a $15$-node network is saturated even at $\lambda{=}9$.
Therefore, network size plays an important role in attaining saturation level of a network, its impact being much higher than even packet arrival rate.
Since our model matches very closely with simulation results (as shown above),
in the rest of this paper we present analysis based on theoretical data derived
from the model.

Fig.~\ref{fig:math}\subref{fig:A1} presents the throughput and channel access delay as a function
of initial contention window $W$ for two packet error rates for the parameters
shown in figure.
Although the most common values of $W$ are $16$ and $32$, a contention window range of $[1{\ldots}100]$ is investigated to demonstrate its effect.
A low $W$ reduces backoff time which in turn increases collisions and decreases
throughput.
However, decrease in the idle period reduces overall channel access time.
Although number of stations and packet arrival rate are kept constant, $d_C$
is initially low but grows quickly
until $W{=}20$, e.g., $d_C{=}0.054$s and $0.065$s at $W{=}10$ and $20$ for $p_E{=}0$.
Increase in channel access delay becomes minimal for $W{>}70$.
Throughput shows a similar trend but apparently the effect of channel error
on throughput is
more aggressive compared to that of $W$.

Fig.~\ref{fig:math}\subref{fig:A3A} shows throughput and channel access delay for a range of retry limits in presence of channel errors.
Delay is found to be lower for a lower $m$.
When $m{=}0$, a packet will be dropped even after the first failure in its transmission.
With higher $m$, a higher number of failed transmissions attempts are tolerated.
Therefore, the HoQ packet can hog the channel for a longer period and channel access delay becomes higher.
On the contrary to $\infty$-limit BEB, where delay upper limit is high, $m$-retry
BEB keeps the channel access delay within a defined margin.
In delay sensitive applications (e.g., VoIP, video conference, etc.), a late packet is as good as a dropped packet.
Playing a late packet will only increase jitter at the receiver end.
Therefore, it is better to drop a late packet when it can help other packets to reach in time~\cite{Siddique2008}.
For the parameters shown in the figure, $d_C$ is found to be $66.9$ms for $m{=}0$, which reaches $67.2$ms for $m{=}3$ and
remains unchanged for higher values of $m$.
On the other hand, throughput is higher for lower $m$ (e.g., $0.6279$ mbps
for $m{=}1$), drops exponentially
with increasing $m$, and becomes nearly constant for $m{\geq}4$.

Fig.~\ref{fig:math}\subref{fig:A4A} elaborates throughput and channel access delay as a function of the number of stations for two packet arrival rates.
With $\lambda{=}2$, throughput is 260.027 kbps at $n{=}1$ which rises very quickly to 892.397 kbps at $n{=}5$ and remains unchanged for higher number
of nodes.
We omit discussion on higher number of stations since the network would then operate in saturated region while we mainly investigate the unsaturated condition in this
paper.
Interested readers may consult \cite{Chatzimisios2004} and the references therein for more discussions on the saturated condition.
With $\lambda{=}6$ the throughput is much higher initially (699.483 kbps at $n{=}1$) and quickly reaches 891.681 kbps at $n{=}3$.
As $n$ increases, throughput remains the same irrespective of $\lambda$.
For $\lambda{=}2$, channel access delay is initially very low ($d_C{=}462\mu s$) at $n{=}1$ which increases slowly at the beginning until $n{=}3$ ($d_C{=}200 \mu s$). Apparently, the offered network is too small to provide sufficient contention. After that, it increases almost linearly with $n$.

A steady state system is defined by transmission probability and collision probability. In Fig.~\ref{fig:queue}\subref{fig:B1}$\sim$~\ref{fig:queue}\subref{fig:B7}, collision probability $p_C$, transmission probability $\tau$, non-empty queue probability $p_Q$, network loss $e_N$, and queue loss $e_Q$ are shown for varying packet arrival rate, initial collision window, and retry limit.
With increase in packet arrival rate, as shown in Fig.~\ref{fig:queue}\subref{fig:B1}, the probability that the queue is non-empty rises very quickly and becomes
$1$ for $\lambda{>}20$.
With the queue always non-empty, further increase in packet arrival rate increases the number of packets in the queue.
When a newly generated packet finds the queue full, it will be dropped.
Therefore, soon queue loss will overwhelm the system and loss becomes very high.
That is why queuing loss $e_Q$ starts growing very quickly from $\lambda{=}20$.
But notably transmission probability $\tau$ remains unchanged after some initial increase up to $\lambda{=}20$.
Since the number of stations is constant, increase in $\lambda$ does not have additional impact once the nodes become nearly saturated.

Fig.~\ref{fig:queue}\subref{fig:B4}, on the other hand, elaborates the same performance measures for different values of contention window $W$.
A higher contention window gives greater spread of counter values over contenders ensuring less collision.
At the same time, nodes will wait longer in idle state and channel access time increases.
We find transmission probability to decrease with increasing collision window.
Probability of a non-empty queue $p_Q$ rises quickly and becomes $1$ as before but queuing loss is much smaller in this case.
Since the packet arrival rate and the number of stations are constant, queuing loss
$e_Q$ does not increase after reaching its highest value of 0.2931.
Transmission probability, collision probability, and network loss decrease
initially with increase in $W$ since nodes tend to spend more time idly and frequency of transmissions become lower.

Fig.~\ref{fig:queue}\subref{fig:B7} presents the above mentioned parameters for variation of retry limit.
At a lower $m$, a packet will be dropped after a smaller number of failures.
As a result, number of transmissions will be less and network loss will be higher.
However, it also means that channel access delay will have a lower upper-limit.
We find that transmission probability decreases initially and remains constant when the network configuration can transmit the offered load.
Collision probability and queuing loss follow a similar pattern of different magnitudes.
Since network loss is given by $p^{m+1}_F$, $e_N$ continues to decrease exponentially with increasing $m$ in the constant $p_C$ region.
Probability of the queue being non-empty is $1$ when packet arrival rate is greater than or equal to servicing rate of the queue i.e., $\lambda{\geqslant} \frac{1}{d_C}$ or $\lambda d_C {\geqslant} 1$.
Channel access delay decreases with increase in $m$ as discussed.
In particular, $d_C{=}0.209$ at $m{=}1$ and $d_C{=}0.0707$ at $m{=}10$.
Therefore, albeit $\lambda$ (${=}20$) is kept constant, $\lambda d_C {\geqslant} 1$ for the whole region and $p_Q{=}1$ for this configuration.

\section{Conclusion}
\label{section:Conclusion}

This paper introduces a Markov model analysis for IEEE 802.11 $m$-retry BEB
based DCF under unsaturated condition.
We investigated
the most widely used network performance measures in an imperfect channel
with provision for power capture.
The performance measures are presented in terms of
network parameters and will be of great assistance in designing and
assessing IEEE 802.11 based wireless networks.
WLANs tend to suffer from severe bottlenecks
formed around the AP under high load which can cause call drop, call
rejection~\cite{Siddique2008} for voice/video calls, and degradation of
VoIP voice quality.
Therefore, these networks should be carefully designed considering
expected load and network parameters.
This model considers both network load and configuration, and is equally
applicable for both saturated and unsaturated studies and, therefore,
can play a significant role in network planning.

\bibliographystyle{IEEEtran}
\bibliography{IEEEabrv,GC09}

\begin{thebibliography}{10}
\providecommand{\url}[1]{#1}
\csname url@samestyle\endcsname
\providecommand{\newblock}{\relax}
\providecommand{\bibinfo}[2]{#2}
\providecommand{\BIBentrySTDinterwordspacing}{\spaceskip=0pt\relax}
\providecommand{\BIBentryALTinterwordstretchfactor}{4}
\providecommand{\BIBentryALTinterwordspacing}{\spaceskip=\fontdimen2\font plus
\BIBentryALTinterwordstretchfactor\fontdimen3\font minus
  \fontdimen4\font\relax}
\providecommand{\BIBforeignlanguage}[2]{{%
\expandafter\ifx\csname l@#1\endcsname\relax
\typeout{** WARNING: IEEEtran.bst: No hyphenation pattern has been}%
\typeout{** loaded for the language `#1'. Using the pattern for}%
\typeout{** the default language instead.}%
\else
\language=\csname l@#1\endcsname
\fi
#2}}
\providecommand{\BIBdecl}{\relax}
\BIBdecl

\bibitem{Bianchi2000}
G.~Bianchi, ``Performance analysis of the {IEEE} 802.11 distributed
  coordination function,'' \emph{{IEEE} J. Sel. Areas Commun.}, pp. 535--547,
  2000.

\bibitem{Ziouva2002}
E.~Ziouva and T.~Antonakopoulos, ``{CSMA\/CA} performance under high traffic
  conditions: throughput and delay analysis,'' \emph{Computer Communications},
  pp. 313--321, 2002.

\bibitem{Wu2002}
H.~Wu, Y.~Peng, K.~Long, S.~Cheng, and J.~Ma, ``{P}erformance of reliable
  transport protocol over {IEEE} 802.11 wireless {LAN}: analysis and
  enhancement,'' in \emph{Proc. IEEE INFOCOM}, 2002, pp. 599--607.

\bibitem{Chatzimisios2005}
P.~Chatzimisios, A.~C. Boucouvalas, and V.~Vitsas, ``Performance analysis of
  the {IEEE} 802.11 {MAC} protocol for wireless {LANs},'' \emph{International
  Journal of Communication Systems}, vol.~18, pp. 545--569, 2005.

\bibitem{Chatzimisios2004}
------, ``{P}erformance analysis of {IEEE} 802.11 {DCF} in presence of
  transmission errors,'' in \emph{Proc. IEEE ICC}, 2004, pp. 3854--3858.

\bibitem{Daneshgaran2008}
F.~Daneshgaran, M.~Laddomada, F.~Mesiti, and M.~Mondin, ``{U}nsaturated
  throughput analysis of {IEEE} 802.11 in presence of non ideal transmission
  channel and capture effects,'' \emph{{IEEE} Trans. Wireless Commun.}, pp.
  1276--1286, 2008.

\bibitem{Barowski2005}
Y.~Barowski, S.~Biaz, and P.~Agrawal, ``Towards the performance analysis of
  {IEEE} 802.11 in multi-hop ad-hoc networks,'' in \emph{Proc. IEEE WCNC},
  2005, pp. 100--106.

\bibitem{Liaw2005}
Y.~S. Liaw, A.~Dadej, and A.~Jayasuriya, ``{P}erformance analysis of {IEEE}
  802.11 {DCF} under limited load,'' in \emph{Proc. Asia Pacific Conference on
  Communications}, 2005, pp. 759--763.

\bibitem{Siddique2008}
M.~A.~R. Siddique and J.~Kamruzzaman, ``{V}o{IP} call capacity over wireless
  mesh networks,'' in \emph{Proc. IEEE GLOBECOM}, 2008, pp. 1--6.

\bibitem{Siddique2009}
------, ``{VoIP} capacity over {PCF} with imperfect channel,'' in \emph{Proc.
  IEEE GLOBECOM to appear}, 2009, pp. 1--6.

\bibitem{Hadzi-velkov2002}
Z.~Hadzi-velkov and B.~Spasenovski, ``Capture effect in ieee 802.11 basic
  service area under influence of rayleigh fading and near/far effect,'' in
  \emph{Proc. IEEE PIMRC}, 2002, pp. 172--176.

\end{thebibliography}
\end{document}